\documentclass[lettersize,journal]{IEEEtran}
\usepackage{amsmath,amsfonts}
\usepackage{algorithmic}
\usepackage{algorithm}
\usepackage{array}
\usepackage[caption=false,font=normalsize,labelfont=sf,textfont=sf]{subfig}
\usepackage{textcomp}
\usepackage{stfloats}
\usepackage{url}
\usepackage{verbatim}
\usepackage{graphicx}
\usepackage{cite}
\usepackage[colorlinks=true, allcolors=blue]{hyperref}
\usepackage{orcidlink}
\usepackage{svg}
\usepackage{bm}
\usepackage{booktabs}

\begin{document}

\title{SpatialNet with Binaural Loss Function for Correcting Binaural Signal Matching Outputs under Head Rotations\\
}

\author{
    Dor Shamay, Boaz Rafaely \\[0.7em]
    \textit{School of Electrical and Computer Engineering, Ben-Gurion University of the Negev}
}

\maketitle

\begin{abstract}
Binaural reproduction is gaining increasing attention with the rise of devices such as virtual reality headsets, smart glasses, and head-tracked headphones. Achieving accurate binaural signals with these systems is challenging, as they often employ arbitrary microphone arrays with limited spatial resolution. The Binaural Signals Matching with Magnitude Least-Squares (BSM-MagLS) method was developed to address limitations of earlier BSM formulations, improving reproduction at high frequencies and under head rotation. However, its accuracy still degrades as head rotation increases, resulting in spatial and timbral artifacts, particularly when the virtual listener’s ear moves farther from the nearest microphones. In this work, we propose the integration of deep learning with BSM-MagLS to mitigate these degradations. A post-processing framework based on the SpatialNet network is employed, leveraging its ability to process spatial information effectively and guided by both signal-level loss and a perceptually motivated binaural loss derived from a theoretical model of human binaural hearing. The effectiveness of the approach is investigated in a simulation study with a six-microphone semicircular array, showing its ability to perform robustly across head rotations. These findings are further studied in a listening experiment across different reverberant acoustic environments and head rotations, demonstrating that the proposed framework effectively mitigates BSM-MagLS degradations and provides robust correction across substantial head rotations.
\end{abstract}


\section{Introduction}
High-quality binaural audio reproduction is essential for recreating immersive acoustic environments, enabling users to perceive sound as if they were physically present. This realism is vital for applications such as virtual and augmented reality and teleconferencing \cite{ref1}. Achieving such immersion requires precise capture and reproduction of the sound field, along with the application of head-related transfer functions (HRTFs) that account for the effect of the head. Traditionally, spatial audio formats like Ambisonics have been highly effective in the reproduction of binaural signals by representing sound fields through spherical harmonics, allowing for accurate three-dimensional spatial representation \cite{ref2}. However, the encoding of Ambisonics often necessitates specific microphone configurations, such as spherical arrays, which may not be practical for applications like wearable or mobile devices \cite{ref1}. As a result, research has increasingly focused on developing flexible approaches that can work with less specialized arrays while delivering high-quality spatial audio.

A recently developed approach for binaural signal reproduction with general array configurations is Binaural Signal Matching (BSM), which estimates binaural signals by minimizing the mean-squared error (MSE) between array steering vectors and HRTFs. This method is signal-independent, as it does not rely on detailed sound-field parameters or source localization. Instead, it prioritizes consistent spatial audio reproduction across a diverse range of sound fields. A recent adaptation within the BSM framework is BSM-MagLS \cite{ref3}. This method provides a robust design framework that effectively accommodates various array geometries and establishes explicit conditions for accurate binaural reproduction, thereby extending its applicability to arbitrary sound fields. Additionally, it employs Magnitude Least-Squares (MagLS) \cite{ref4} instead of Least-Squares (LS) at high frequencies, where the accuracy of BSM tends to considerably degrade. BSM-MagLS has proven effective in enhancing high-frequency performance and demonstrates utility in compensating for listener head rotations during playback. However, it still faces challenges when the ears of a virtual listener move farther from the nearest microphones in the array under head rotations \cite{ref5}. Consequently, spatial and timbral artifacts become more pronounced in the ear farther from the array, with the severity of these artifacts increasing as the distance between the virtual ear and the microphones grows.

One possible way to address the degradation of BSM-MagLS accuracy during head rotations, is using signal-dependent approaches, such as those in \cite{ref6,ref7}, that may offer a solution by leveraging spatial information in the sound field. These methods estimate important parameters, including the number of sources and their directions of arrival (DOAs), prior to playback, enabling accurate spatial rendering despite head movements. However, they rely heavily on precise parameter estimation, where errors can introduce artifacts and degrade sound quality, limiting their applicability in complex acoustic environments. Further refinement of estimation techniques within signal-dependent frameworks remains an important research direction and falls outside the scope of this study.

In this paper, we propose the integration of deep learning with BSM-MagLS to address the challenges posed by head rotations. Our approach employs a post-processing framework where the SpatialNet neural network \cite{ref16} is applied to the output of BSM-MagLS to correct distortions introduced during head-rotation compensation. By leveraging the more reliable signal estimation from the ear closer to the microphone array, the network adjusts the signals at both ears, contributing to the preservation of spatial cues, reduction of timbral distortions, and improvement of spatial accuracy. 
Complementing the post-processing approach, we also address the design of loss functions. Prior studies in binaural audio have predominantly relied on loss functions defined in the short-time Fourier transform (STFT) domain, focusing on interaural level difference (ILD) and interaural phase difference (IPD) cues \cite{ref8,ref9}. 
In this work, we instead employ a binaural loss derived from a model of human binaural hearing \cite{ref10}, which is based on auditory filters and has not previously been applied as a loss function for training neural networks. We evaluate the proposed method and assess the effectiveness of the binaural loss through both objective metrics and a subjective listening experiment. Together, these evaluations provide a comprehensive validation of the approach in mitigating distortions that arise in BSM-MagLS under head rotation.

\section{Binaural Signal Matching with Magnitude Least Squares (BSM-MagLS)}

This section provides a background for the BSM-MagLS binaural reproduction approach \cite{ref3}, introducing fundamental concepts necessary for understanding its framework. The analysis throughout this section adopts the spherical coordinate system, denoted by $(r, \theta, \phi)$, where $r$ is the radial distance from the origin, $\theta$ is the elevation angle measured downward from the Cartesian $z$ axis to the $xy$ plane, and $\phi$ represents the azimuthal angle, defined from the positive $x$ axis toward the positive $y$ axis.

\subsection{Microphone Array and Binaural Signal Models  }

Assume a sound field composed of \( D \) far-field sound sources, emitting the signals \(\{s_d(k)\}_{d=1}^{D}\) with directions-of-arrival (DOAs) \(\{(\theta_d, \phi_d)\}_{d=1}^{D}\). Here, \( k = \frac{2\pi}{\lambda} \) is the wave-number, and \( \lambda \) is the wavelength of the wave. The sound field is captured by a microphone array consisting of \( M \) elements, centrally positioned at the origin, with microphone locations given by \( \{(r_m, \theta_m, \phi_m)\}_{m=1}^{M} \). The noisy measurements captured by the array can be described using the following narrowband representation \cite{ref11}:
\begin{equation}
    \mathbf{x}(k) = \mathbf{V}(k)\mathbf{s}(k) + \mathbf{n}(k)
    \label{eq:1}
\end{equation}
where \(\mathbf{x}(k) = \begin{bmatrix} x_1(k) , \cdots , x_M(k) \end{bmatrix}^T\) is the vector of the pressure amplitudes measured by the microphones, \(\mathbf{V}(k) = \begin{bmatrix} \mathbf{v}_1(k, \theta_1, \phi_1) , \cdots , \mathbf{v}_D(k, \theta_D, \phi_D) \end{bmatrix}\) is an \(M \times D\) steering matrix with columns \(\mathbf{v}_d(k) = \begin{bmatrix} v_1(k, \theta_d, \phi_d) , \cdots , v_M(k, \theta_d, \phi_d) \end{bmatrix}^T\), where each element represents the transfer function from source \(d\) to microphone \(m\), \(\mathbf{s}(k) = \begin{bmatrix} s_1(k) , \cdots , s_D(k) \end{bmatrix}^T\) is the source signals vector, \(\mathbf{n}(k) = \begin{bmatrix} n_1(k) , \cdots , n_M(k) \end{bmatrix}^T\) is the additive noise vector, and \((\cdot)^T\) is the transpose operator.

Assuming the listener's head is centered at the origin, and that the sound field comprises \( D \) far-field sound sources, the pressure at the ears can be expressed as \cite{ref12}:  
\begin{equation}
    p^{l,r}(k) = [\mathbf{h}^{l,r}(k)]^T \mathbf{s}(k),
    \label{eq:2}
\end{equation}
where \( p^{l,r}(k) \) are the complex-valued sound pressures, and  
\( \mathbf{h}^{l,r}(k) = \begin{bmatrix} h^{l,r}(k,\theta_1,\phi_1), \dots, h^{l,r}(k,\theta_D,\phi_D) \end{bmatrix}^T \) are the vectors containing the HRTFs corresponding to the DOAs of the sources for the left and right ears, denoted by \( (\cdot)^l \) and \( (\cdot)^r \), respectively.

\subsection{Computing BSM filters}

The initial step in the BSM approach involves linearly combining the microphone signals through spatial filtering, represented as:
\begin{equation}
    z^{l,r}(k) = [\mathbf{c}^{l,r}(k)]^H \mathbf{x}(k),
    \label{eq:3}
\end{equation}
where \( \mathbf{c}^{l,r}(k) = \begin{bmatrix} c^{l,r}_1(k), \dots, c^{l,r}_M(k) \end{bmatrix}^T \) is the complex-valued vector containing the filter coefficients, and \( (\cdot)^H \) is the Hermitian transpose operator. 

Subsequently, the vectors \( \mathbf{c}^{l,r} \) are optimized to minimize the mean-squared error between the filtered signals and the corresponding binaural pressure signals in \eqref{eq:2}, for each ear independently:
\begin{equation}
    \epsilon^{l,r}(k) = \mathbb{E}\left[ | p^{l,r}(k) - z^{l,r}(k) |^2 \right],
    \label{eq:4}
\end{equation}
where \( \mathbb{E}[\cdot] \) is the expectation operator. 

To reduce the dependence on specific information about the sound field, it is assumed that the sources are uncorrelated with equal power \( \sigma_s^2 \), i.e. a diffuse sound field, and that the noise is both white and uncorrelated with the sources as well as between microphones, with power \( \sigma_n^2 \). Based on these assumptions, the MSE is expressed as:
\begin{equation}
    \epsilon^{l,r}(k) = \sigma_s^2 \left\| \mathbf{V}^T(k) [\mathbf{c}^{l,r}(k)]^* - \mathbf{h}^{l,r}(k) \right\|_2^2 + \sigma_n^2 \left\| [\mathbf{c}^{l,r}(k)]^* \right\|_2^2,
    \label{eq:5}
\end{equation}
where \( \|\cdot\|_2^2 \) is the \( l^2 \)-norm. 

The solution that minimizes this error is then given by \cite{ref13}:
\begin{equation}
    \mathbf{c}_{\text{BSM}}^{l,r}(k) =
    \bigl( \mathbf{V}(k)\mathbf{V}^H(k) + \tfrac{\sigma_n^2}{\sigma_s^2}\mathbf{I}_M \bigr)^{-1}
    \mathbf{V}(k)\,[\mathbf{h}^{l,r}(k)]^*,
    \label{eq:6}
\end{equation}
where \( \mathbf{I}_M \) is the identity matrix of size \( M \), and \( (\cdot)^* \) is the complex conjugate operator. 

Finally, by substituting \eqref{eq:6} into \eqref{eq:3}, the estimated binaural signals using the BSM approach, denoted as \( \hat{p}^{l,r}(k) \), are obtained:
\begin{equation}
    \hat{p}_{\text{BSM}}^{l,r}(k) = [\mathbf{c}_{\text{BSM}}^{l,r}(k)]^H \mathbf{x}(k).
    \label{eq:7}
\end{equation}

The BSM framework was shown in \cite{ref3} to be valid also for arbitrary sound fields, when \( D \) is sufficiently large and the error in \eqref{eq:5} remains sufficiently small.

\subsection{BSM with Magnitude Least Squares}

It was shown in \cite{ref3} that under practical conditions, BSM performance may degrade, particularly at high frequencies. To address this, the principle that IPD becomes perceptually insignificant at high frequencies, typically above 1.5 kHz \cite{ref14,ref15}, is applied. Thus, the solution in \eqref{eq:6} up to this cutoff is combined with the coefficients derived from the following minimization above the cutoff, where the traditional MSE formulation is replaced by matching absolute values:
\begin{equation}
\begin{aligned}
    \mathbf{c}_{\text{BSM-MagLS}}^{l,r}(k) 
    &= \arg \min_{\mathbf{c}^{l,r}(k)} \, 
    \Big[ \sigma_s^2 
    \big\| | \mathbf{V}^T(k) [\mathbf{c}^{l,r}(k)]^* | 
    - | \mathbf{h}^{l,r}(k) | \big\|_2^2 \\
    &\quad + \sigma_n^2 
    \big\| [\mathbf{c}^{l,r}(k)]^* \big\|_2^2 \Big].
\end{aligned}
\label{eq:8}
\end{equation}

This solution, referred to as magnitude least-squares (MagLS), leads to reduced magnitude errors.

\subsection{Compensating for Listener's Head Rotations}
During the playback stage of BSM reproduction, it is desired to reproduce binaural signals in a world-locked acoustic environment relative to the listener. Therefore, head rotations must be compensated for. It is assumed that the exact degree of head rotation is known via a head-tracking device, with the restriction that head rotations are confined to the azimuthal plane for simplicity. The array and its orientation during acoustic scene recording, along with the initial alignment of the listener's head, are illustrated in Fig.~\hyperref[fig:1]{\ref*{fig:1}(a)}. During playback, the orientation of the listener’s head becomes relevant, as it may rotate relative to the array and form a relative angle.
This is illustrated in Fig.~\hyperref[fig:1]{\ref*{fig:1}(b)}, where a head rotation of \( \phi_{\text{rot}} \) degrees is applied. To compensate for this head rotation, the HRTF vectors \( \mathbf{h}^{l,r}(k) \) used in the calculation of the filters \( \mathbf{c}_{\text{BSM}}^{l,r} \) and \( \mathbf{c}_{\text{BSM-MagLS}}^{l,r} \) in \eqref{eq:6} and \eqref{eq:8} are modified to represent the HRTFs rotated by \( \phi_{\text{rot}} \) degrees relative to the array’s original position. The modified HRTF vectors, denoted as \( \mathbf{h}_{\text{rot}}^{l,r}(k) \), are given by:
\begin{equation}
    \mathbf{h}_{\text{rot}}^{l,r}(k) =
    \begin{bmatrix} 
        h^{l,r}(k, \theta_1, \phi_1 + \phi_{\text{rot}}),\! \dots,\! 
        h^{l,r}(k, \theta_D, \phi_D + \phi_{\text{rot}})
    \end{bmatrix}^T.
    \label{eq:9}
\end{equation}

\subsection{Performance with Head Rotation Compensation}
The BSM-MagLS method is designed to accurately reproduce binaural signals for a sound field captured by a microphone array with \( M \) elements. However, the effectiveness of head rotation compensation diminishes as the distance between the microphones and the ear positions in the HRTFs increases \cite{ref5}. This degradation is particularly pronounced in array configurations where one ear is positioned farther from the microphones under rotation, as in semi-circular arrays. In such configurations, the degree of head rotation exacerbates the degradation, causing the quality of the signal reproduced at the ear farther from the microphone array to deteriorate, while the signal at the ear closer to the microphones remains relatively accurate. Addressing this degradation is the focus of the present work, as detailed in the following section.

\section{SpatialNet for Correcting BSM-MagLS Outputs with Head Rotations}
In this section, we present an overview of the proposed method and of SpatialNet \cite{ref16}, which aim to correct the BSM-MagLS binaural output under head rotation.

\subsection{Overview of the proposed method}\label{AA}
\label{sec:3.1}
\begin{figure*}[!t] 
\centering
\includegraphics[width=\textwidth]{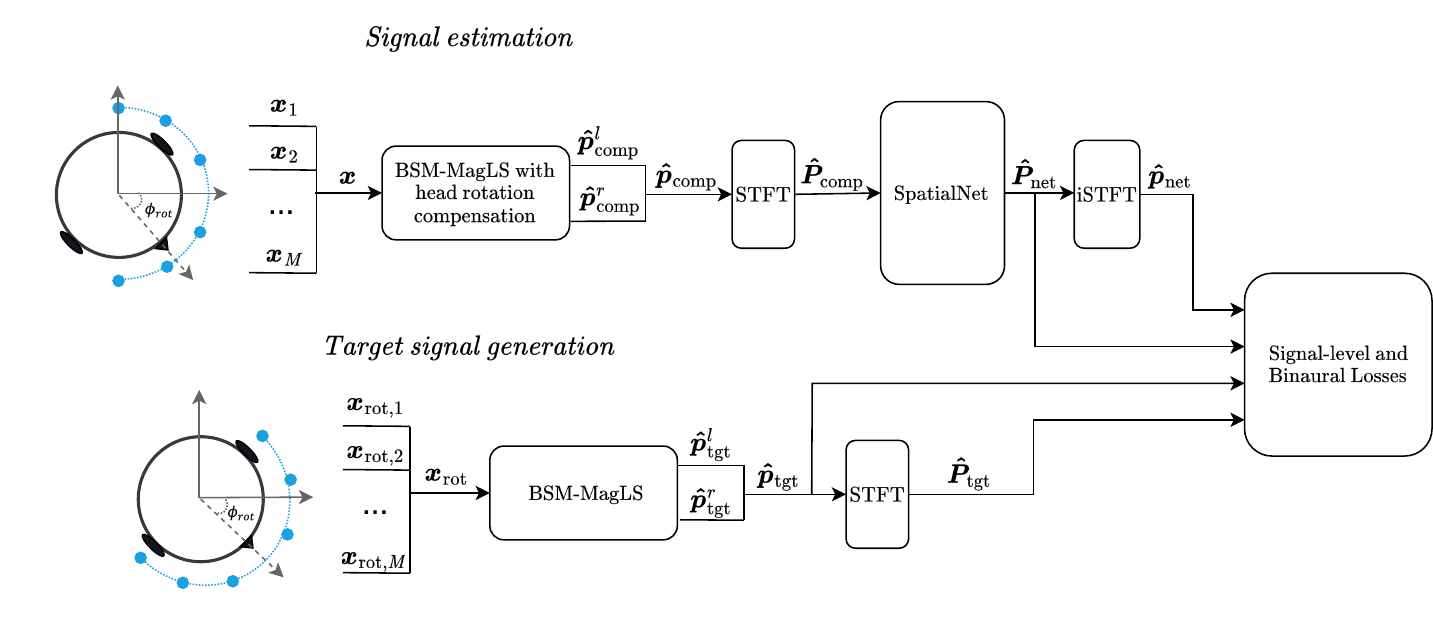}
\vskip -3ex 
\caption{Schematic overview of the system with signal-level and binaural losses.}
\label{fig:2}
\end{figure*}

A schematic overview of the proposed system is presented in Fig.\hyperref[fig:2]{~\ref*{fig:2}}. The upper pipeline represents the signal estimation, starting with applying BSM-MagLS to the microphone array recording in its original position, with head-rotation compensation applied for the head rotation assumed during reproduction. The time-domain compensated binaural output, $\hat{\bm{p}}_{\text{comp}}^{l,r}$, is then processed using STFT and fed as input to the SpatialNet model, which yields the STFT of the estimated binaural signal. The time-domain signal, $\hat{\bm{p}}_{\text{net}}^{l,r}$, is then obtained by applying the inverse STFT (iSTFT). 

The lower pipeline represents the target signal generation, where the microphone array is rotated to match the listener's head orientation, according to the assumed rotation during reproduction. This choice of target signal uses the same settings as in the upper pipeline, with the array aligned to the head orientation. In this configuration no compensation is required, and BSM-MagLS provides an accurate estimation that serves as a reliable reference for correcting errors due to head rotation. The rotated microphone array recording, $\bm{x}_{\text{rot}}$, is then processed using BSM-MagLS to yield the target binaural signal, $\hat{\bm{p}}_{\text{tgt}}^{l,r}$.

\begin{figure}[!t] 
\centering
\includegraphics[width=\linewidth]{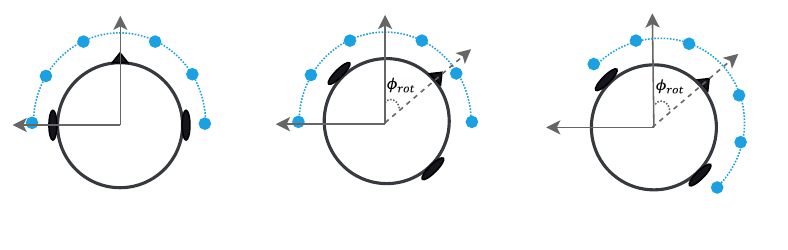}
\vskip -3ex 
{\footnotesize \hspace{0.45cm}
(a) Original \hspace{1.1cm}
(b) Head rotation \hspace{0.35cm}
(c) Array and head rotation
}
\caption{Schematic diagram of a listener's head and a semi-circular array with $M=6$ microphones (blue dots). (a) Original orientation, (b) head rotation by $\phi_{\text{rot}}$ degrees in the azimuth, and (c) rotation of both array and the head by $\phi_{\text{rot}}$ degrees in the azimuth.}
\label{fig:1}
\end{figure}

In practice, to form the target signal, the positions of the microphones are adjusted to $\{(r_m, \theta_m, \phi_m + \phi_{\text{rot}})\}_{m=1}^M$. In this scenario, depicted in Fig.~\hyperref[fig:1]{\ref*{fig:1}(c)}, the BSM-MagLS spatial filter coefficients are computed by minimizing a term dependent on a rotated steering matrix, but with the same HRTF coefficients as in the case of head rotation compensation, since the head orientation is identical in both cases.

The proposed method is based on SpatialNet, which is designed to effectively process spatial information, relevant in this study incorporating binaural signals. The network leverages both narrow-band and cross-band information, including residual connections, which are particularly beneficial as they help preserve the underlying structure of the input binaural signal. Moreover, the time-convolutional module in the narrow-band block may be well-suited for correcting distortions arising from the convolutive effects of inaccurate filters produced by the BSM-MagLS method in the case of head rotation.

During training, the estimated and target binaural signals, along with their corresponding STFTs, are compared using signal-level and binaural loss terms. These losses guide the refinement of the SpatialNet model weights in estimating the target signal.

\subsection{SpatialNet}

\begin{figure}[!t] 
\centering
\includegraphics[width=\linewidth]{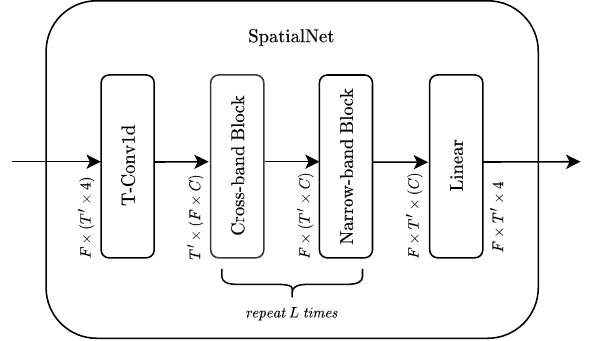}
\caption{SpatialNet architecture overview.}
\label{fig:3}
\end{figure}

The SpatialNet architecture was originally proposed as an efficient and high-quality model for multi-channel to single-channel speech separation, denoising, and dereverberation. A systematic overview of the SpatialNet architecture as applied in this work is shown in Fig.~\hyperref[fig:3]{\ref*{fig:3}}. In contrast to the original formulation, which maps multi-channel array input to separated speech sources, our adaptation takes degraded two-channel binaural signals as input and predicts corrected binaural audio as output.

The network operates in the time-frequency (T-F) domain by applying the STFT to the time-domain binaural waveform $\hat{\bm{p}}_{\text{comp}} = [\hat{\bm{p}}_{\text{comp}}^l, \hat{\bm{p}}_{\text{comp}}^r] \in \mathbb{R}^{T \times 2}$, where $T$ denotes the number of samples. The input to the network consists of the STFT coefficients, denoted $\hat{\bm{P}}_{\text{comp}}$, where the real and imaginary parts of each binaural channel are concatenated for each T-F bin:
\[
\begin{aligned}
\bm{\hat{P}}_{\text{comp}}[f,t',:] = [ \,
    & \mathcal{R}(\hat{P}_{\text{comp}}^l(f,t')), \,
    \mathcal{I}(\hat{P}_{\text{comp}}^l(f,t')), \\
    & \mathcal{R}(\hat{P}_{\text{comp}}^r(f,t')), \,
    \mathcal{I}(\hat{P}_{\text{comp}}^r(f,t')) \,
] \in \mathbb{R}^4.
\end{aligned}
\]
where $\mathcal{R}(\cdot)$ and $\mathcal{I}(\cdot)$ denote the real and imaginary parts, respectively, and $t' \in \{1, \dots, T'\}$, $f \in \{0, \dots, F-1\}$ represent the time-frame and frequency-bin indices. The network output, denoted $\hat{\bm{P}}_{\text{net}}$, corresponds to the predicted concatenated STFT coefficients of the target binaural signal, $\hat{\bm{P}}_{\text{tgt}}$, for each T-F bin:
\[
\begin{aligned}
\bm{\hat{P}}_{\text{tgt}}[f,t',:] = [ \,
    & \mathcal{R}(\hat{P}_{\text{tgt}}^l(f,t')), \,
    \mathcal{I}(\hat{P}_{\text{tgt}}^l(f,t')), \\
    & \mathcal{R}(\hat{P}_{\text{tgt}}^r(f,t')), \,
    \mathcal{I}(\hat{P}_{\text{tgt}}^r(f,t')) \,
] \in \mathbb{R}^4.
\end{aligned}
\]

As shown in Fig.~\hyperref[fig:3]{\ref*{fig:3}}, SpatialNet consists of a convolutional input layer (T-Conv1d), $L$ interleaved narrow-band and cross-band blocks, and a linear output layer. The convolutional input layer applies a convolution to $\bm{\hat{P}}_{\text{comp}}$, producing a hidden representation of shape $\mathbb{R}^{F \times T' \times C}$, where the kernel operates along the time dimension. This representation is then processed by the interleaved narrow-band and cross-band blocks, with all modules within each block connected via residual connections. The linear output layer maps the output of the final block to the predicted concatenated STFT coefficients $\hat{\bm{P}}_{\text{net}}$. Finally, the time-domain binaural signal is obtained by applying the iSTFT, resulting in $\hat{\bm{p}}_{\text{net}}$, which represents the estimation of the time-domain target binaural signal $\hat{\bm{p}}_{\text{tgt}} = [\hat{\bm{p}}_{\text{tgt}}^l,\, \hat{\bm{p}}_{\text{tgt}}^r] \in \mathbb{R}^{T \times 2}$.

A detailed description of the modules within each block is provided in \cite{ref16}. The narrow-band block is designed to process each frequency bin independently, leveraging the well-established spatial features inherent in narrow-band signals. It consists of a multi-head self-attention (MHSA) module and a time-convolutional feedforward network (T-ConvFFN). This block operates on individual STFT frequency bins, with all frequencies sharing the same network parameters, achieved by folding the frequency axis into the batch dimension during training. The cross-band block contains two frequency-convolutional layers and one full-band linear module. It operates independently on each time frame, with the time axis similarly folded into the batch dimension, sharing weights across time frames. The frequency-convolutional module models correlations between adjacent frequency bins, while the full-band linear module addresses the limitations of narrow-band modeling by incorporating spatial dependencies across the entire frequency spectrum.

\section{Loss Function}
In contrast to previous works that have focused on loss functions directly defined for interaural cues \cite{ref8,ref9}, our approach is tailored to binaural hearing under realistic acoustic conditions \cite{ref10}, \cite{ref17}. This provides a more robust foundation for training, particularly in scenarios where interaural cues, often defined under simplified assumptions, may be less informative.

The proposed loss function consists of a signal-level loss, \( \mathcal{L}_{\text{signal}} \), which focuses on reducing distortion and minimizing time-domain and spectral discrepancies with the target waveform, and a binaural loss, \( \mathcal{L}_{\text{binaural}} \), which aims to preserve the interaural cues essential for accurate spatial perception. The overall loss function is formulated as:
\begin{equation}
    \mathcal{L} = \mathcal{L}_{\text{signal}} + \mathcal{L}_{\text{binaural}}.
    \label{eq:10}
\end{equation}

The remainder of this section introduces the signal-level losses, outlines the auditory processing used for binaural cue extraction, defines the proposed auditory-inspired binaural loss, and finally presents an alternative STFT-based binaural loss for comparison.

\subsection{Signal-Level Loss}
The components of the signal-level loss are defined as follows. The scale-invariant signal-to-distortion ratio (\text{SI-SDR})~\cite{ref18} loss, denoted \( \mathcal{L}_{\text{SI-SDR}} \), is computed for the binaural output signal \( \hat{\bm{p}}_{\text{net}} \) with respect to the target signal  \( \hat{\bm{p}}_{\text{tgt}} \):
\begin{equation}
    \mathcal{L}_{\text{SI-SDR}} = -\left( \frac{2}{3} \, \text{SI-SDR}(\hat{\bm{p}}_{\text{tgt}}^r, \hat{\bm{p}}_{\text{net}}^r) + \frac{1}{3} \, \text{SI-SDR}(\hat{\bm{p}}_{\text{tgt}}^l, \hat{\bm{p}}_{\text{net}}^l) \right),
    \label{eq:13}
\end{equation}
where the right channel is assumed, without loss of generality, to correspond to the ear positioned farther from the microphone array during head rotation and is therefore considered less reliable, assigned a correspondingly higher weighting. Information from the more reliable ear is assumed to be readily available given the array and head orientations.
The \text{SI-SDR} is defined as~\cite{ref18}:
\begin{equation}
    \text{SI-SDR}(\hat{\bm{p}}_{\text{tgt}}, \hat{\bm{p}}_{\text{net}}) = 10 \log_{10} \left( \frac{ \left\| \frac{ \hat{\bm{p}}_{\text{net}}^T \hat{\bm{p}}_{\text{tgt}}}{ \| \hat{\bm{p}}_{\text{tgt}} \|^2 } \hat{\bm{p}}_{\text{tgt}} \right\|^2 }{ \left\| \frac{ \hat{\bm{p}}_{\text{net}}^T \hat{\bm{p}}_{\text{tgt}}}{ \| \hat{\bm{p}}_{\text{tgt}} \|^2 } \hat{\bm{p}}_{\text{tgt}} - \hat{\bm{p}}_{\text{net}} \right\|^2 } \right),
    \label{eq:14}
\end{equation}
where \( \| \cdot \| \) denotes the \( \ell_2 \)-norm.

Following the logic of the BSM-MagLS approach, two additional losses are defined according to a perceptually motivated principle. The STFT loss  \( \mathcal{L}_{\text{STFT}} \) is applied to complex coefficients below 1.5 kHz, while the magnitude-based STFT loss \( \mathcal{L}_{\text{Mag-STFT}} \) is applied to coefficients above this threshold.
The loss function \( \mathcal{L}_{\text{STFT}} \) is defined as:
\begin{equation}
    \mathcal{L}_{\text{STFT}} = \| \bm{\hat{P}}_{\text{tgt}} - \bm{\hat{P}}_{\text{net}} \|_1,
    \label{eq:15}
\end{equation}
where \( \| \cdot \|_1 \) denotes the \( \ell_1 \)-norm. The loss function \( \mathcal{L}_{\text{Mag-STFT}} \) is given by:
\begin{equation}
    \mathcal{L}_{\text{Mag-STFT}} = \left\| \log_e | \bm{\hat{P}}_{\text{tgt}} | - \log_e | \bm{\hat{P}}_{\text{net}} | \right\|_1 + \frac{\| | \bm{\hat{P}}_{\text{tgt}} | - | \bm{\hat{P}}_{\text{net}} | \|_{\text{Fro}}}{ \| | \bm{\hat{P}}_{\text{tgt}} | \|_{\text{Fro}} },
    \label{eq:16}
\end{equation}
where the first term represents the logarithmic magnitude loss of the STFT coefficients, and the second term corresponds to the spectral convergence loss. Here, \( | \cdot | \) denotes the absolute value operation, and \( \| \cdot \|_{\text{Fro}} \) represents the Frobenius norm.
The overall signal-level loss combines these components in a weighted sum and is defined as:
\begin{equation}
    \mathcal{L}_{\text{signal}} = \alpha \mathcal{L}_{\text{SI-SDR}} + \beta \mathcal{L}_{\text{STFT}} + \gamma \mathcal{L}_{\text{Mag-STFT}},
    \label{eq:11}
\end{equation}
where the parameters $\alpha$, $\beta$ and $\gamma$ denote the weights applied to each component.

\subsection{Auditory processing for Binaural Loss}
In previous work, interaural cues are extracted from the time-domain binaural signal using an auditory front-end based on the framework presented in \cite{ref10}, which also forms the foundation of the binaural audio quality metric introduced in \cite{ref17}. The auditory processing employed in this work adopts a similar structure, consisting of monaural processing applied independently to the left and right channels, followed by binaural processing to extract interaural cues.

Monaural processing begins by modeling the middle ear transfer characteristics with a first-order bandpass filter in the range 500–2000~Hz.  This is followed by basilar membrane filtering, approximated using a third-order gammatone filter bank comprising 29 frequency bands distributed from 50 to 6000~Hz, with center frequencies spaced one equivalent rectangular bandwidth (ERB) apart. Cochlear compression is applied via instantaneous compression with an exponent of 0.4 to the gammatone filter outputs. Hair cell transduction is then modeled through half-wave rectification followed by a fifth-order low-pass filter with a cutoff frequency of 770~Hz.

In the binaural processing stage, ILD is extracted by applying a second-order Butterworth low-pass filter with a cutoff frequency of 30Hz to the complex monaural outputs of the cochlear compression mechanism. The filtered outputs are denoted as \( a^{l,r}(t,f_{c_k}) \). A de-compression with the exponent is applied, and ILD is calculated as:
\begin{equation}
\mathrm{ILD}(t,f_{c_k}) = \frac{20}{0.4} \log_{10} \left( \frac{|a^r(t,f_{c_k})|}{|a^l(t,f_{c_k})|} \right).
\label{eq:17}
\end{equation}

In a second computation, interaural temporal disparities are derived by applying a second-order gammatone filter to the hair cell transduction monaural outputs, yielding the left and right time-domain complex signals \( g^{l,r}(t,f_c) \) in each frequency band \( f_c \). The interaural transfer function (ITF) was calculated for each band as:
\begin{equation}
\mathrm{ITF}(t,f_c) = g^l(t,f_c) \cdot \overline{g}^r(t,f_c),
\label{eq:18}
\end{equation}
where \( \overline{g}^r \) denotes the complex conjugate of \( g^r \). The IPD is then extracted by computing:
\begin{equation}
\mathrm{IPD}(t,f_c) = \arg \left( \mathrm{ITF}(t,f_c) \right).
\label{eq:19}
\end{equation}
Since sensitivity of the human auditory system to fine temporal structure becomes negligible above approximately 1.4 kHz ~\cite{ref19}, only frequency bands below this limit are used for IPD analysis.

Finally, interaural vector strength (IVS), introduced in \cite{ref10} as a measure of interaural coherence (IC) derived from the ITF in the time domain:
\begin{equation}
\text{IVS}(t, f_{c_k}) = \frac{\left| \int_0^\infty d\tau \, \text{ITF}(t-\tau, f_{c_k}) e^{-\tau / \tau_s} \right|}
{\int_0^\infty d\tau \left| \text{ITF}(t-\tau, f_{c_k}) \right| e^{-\tau / \tau_s}},
\label{eq:20}
\end{equation}
where \( \tau_s \) denotes the integration time constant, set as a multiple of the cycle duration, $T_{\text{c}}$, corresponding to the center frequency of the respective gammatone filter band. A value of \( \tau_s = 5 \cdot T_{\text{c}} \) was found to be optimal for localization performance \cite{ref10}, and was also adopted in the binaural audio quality metric in \cite{ref17}.

\subsection{Binaural Loss Incorporating Auditory Filters}
To leverage ILD, IPD, and IVS as loss functions, we compute the mean squared error (MSE) between each interaural cue of the target and the estimated signal:
\begin{equation}
\mathcal{L}_{\text{ILD}} = \frac{1}{29T} \sum_{t=1}^{T} \sum_{k=1}^{29} \left( \text{ILD}^{\text{tgt}}(t, f_{c_k}) - \text{ILD}^{\text{est}}(t, f_{c_k}) \right)^2,
\label{eq:21}
\end{equation}
\begin{equation}
\mathcal{L}_{\text{IPD}} = \frac{1}{17T} \sum_{t=1}^{T} \sum_{k=1}^{17} \left( \text{IPD}^{\text{tgt}}(t, f_{c_k}) - \text{IPD}^{\text{est}}(t, f_{c_k}) \right)^2,
\label{eq:22}
\end{equation}
\begin{equation}
\mathcal{L}_{\text{IVS}} = \frac{1}{29T} \sum_{t=1}^{T} \sum_{k=1}^{29} \left( \text{IVS}^{\text{tgt}}(t, f_{c_k}) - \text{IVS}^{\text{est}}(t, f_{c_k}) \right)^2,
\label{eq:23}
\end{equation}
where \( \text{ILD}^{\text{tgt}}, \text{IPD}^{\text{tgt}}, \) and \( \text{IVS}^{\text{tgt}} \) denote the target binaural cues, and \( \text{ILD}^{\text{est}}, \text{IPD}^{\text{est}}, \) and \( \text{IVS}^{\text{est}} \) denote the corresponding estimated cues, all computed using (\ref{eq:17}), (\ref{eq:19}), and (\ref{eq:20}), respectively. In these equations, $T$ is the number of time frames and the constants represent the number of frequency bands. The ILD loss \( \mathcal{L}_{\text{ILD}} \), the IPD loss \( \mathcal{L}_{\text{IPD}} \), and the IVS loss \( \mathcal{L}_{\text{IVS}} \) are combined to obtain the overall binaural loss:
\begin{equation}
    \mathcal{L}_{\text{binaural}} = \delta \mathcal{L}_{\text{ILD}} + \lambda \mathcal{L}_{\text{IPD}} + \kappa \mathcal{L}_{\text{IVS}},
    \label{eq:12}
\end{equation}
where $\delta$, $\lambda$ and $\kappa$ denote the weights applied to each interaural cue loss component.

\subsection{Binaural Loss Based on STFT}
As a simpler alternative to the auditory-inspired binaural loss described in (\ref{eq:21})--(\ref{eq:23}), this section presents a binaural loss computed directly from the STFT representations of the binaural signals. A similar loss has been used in prior works, such as in \cite{ref8}. The directly defined ILD and IPD for the target binaural signal are given by:
\begin{equation}
\mathrm{ILD}_{\mathrm{def}}^{\mathrm{tgt}}(f, t') = 20 \log_{10} \left( \frac{|\mathrm{Y}^l(f, t')|}{|\mathrm{Y}^r(f, t')|} \right)
\label{eq:24}
\end{equation}
\begin{equation}
\mathrm{IPD}_{\mathrm{def}}^{\mathrm{tgt}}(f, t') = \arg \left( \frac{\mathrm{Y}^l(f, t')}{\mathrm{Y}^r(f, t')} \right)
\label{eq:25}
\end{equation}
where \( \mathrm{Y}^l \) and \( \mathrm{Y}^r \) are the complex-valued STFT coefficients of the left and right target binaural channels. The ILD and IPD for the estimated binaural signal are computed analogously to (\ref{eq:24}) and (\ref{eq:25}), and are denoted by \( \mathrm{ILD}_{\mathrm{def}}^{\mathrm{net}} \) and \( \mathrm{IPD}_{\mathrm{def}}^{\mathrm{net}} \), respectively. The corresponding loss terms \( \mathcal{L}_{\mathrm{ILD}_{\mathrm{def}}} \) and \( \mathcal{L}_{\mathrm{IPD}_{\mathrm{def}}} \) are computed as:
\begin{equation}
\mathcal{L}_{\mathrm{ILD}_{\mathrm{def}}} = \frac{1}{T' F} \sum_{t'=1}^{T'} \sum_{f=0}^{F-1} \left( \mathrm{ILD}_{\mathrm{def}}^{\mathrm{tgt}}(f, t') - \mathrm{ILD}_{\mathrm{def}}^{\mathrm{net}}(f, t') \right)^2
\label{eq:26}
\end{equation}
\begin{equation}
\mathcal{L}_{\mathrm{IPD}_{\mathrm{def}}} = \frac{1}{T' F} \sum_{t'=1}^{T'} \sum_{f=0}^{F-1} \left( \mathrm{IPD}_{\mathrm{def}}^{\mathrm{tgt}}(f, t') - \mathrm{IPD}_{\mathrm{def}}^{\mathrm{net}}(f, t') \right)^2
\label{eq:27}
\end{equation}
Accordingly, the binaural loss is computed next, as an auditory-inspired alternative to (\ref{eq:12}), defined as:
\begin{equation}
\mathcal{L}_{\mathrm{binaural}_{\mathrm{def}}} = \delta' \mathcal{L}_{\mathrm{ILD}_{\mathrm{def}}} + \lambda' \mathcal{L}_{\mathrm{IPD}_{\mathrm{def}}}
\label{eq:28}
\end{equation}
where $\delta'$ and $\lambda'$ are the STFT-based ILD and IPD loss weights, respectively.

\section{Simulation Study}
\label{sec:5}
In this section, we describe the data generation process, model variants, and training procedures used to develop the proposed method. This is followed by a performance evaluation on simulated scenes with varying acoustic conditions and listener head rotations, using both signal-level and binaural metrics.

\subsection{Setup}
\label{sec:5.1}

The dataset used in this study is based on monaural clean speech signals taken from the LJSpeech dataset~\cite{ref20}, resampled to 16\,kHz and segmented into 30{,}400 two-second excerpts. These speech segments were used as source signals for a simulated point source within a shoebox-shaped room, modeled using the image source method~\cite{ref21} and implemented in MATLAB~\cite{ref22}.

\begin{figure}[!t] 
\centering
\includegraphics[width=0.9\linewidth]{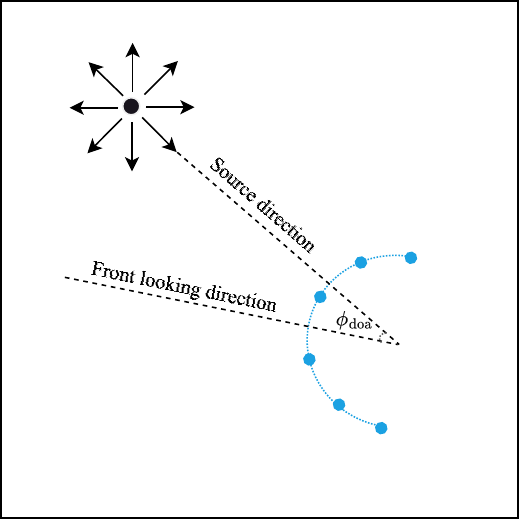}
\caption{Microphone array geometry and definition of DOA angle $\phi_{\mathrm{doa}}$ as the azimuth between the array's front-looking and source directions.}
\label{fig:4}
\end{figure}

A semi-circular microphone array consisting of $M=6$ microphones mounted on a rigid sphere was randomly placed within the simulated room, maintaining a minimum distance of 1\,m from the walls. The room dimensions $(L_x, L_y, L_z)$ were randomly sampled from $L_x, L_y \in [6\,\mathrm{m}, 10\,\mathrm{m}]$ and $L_z \in [3\,\mathrm{m}, 4\,\mathrm{m}]$. Wall reflection coefficients were selected to produce a reverberation time $T_{60}$ that varied randomly from 0.3 to 0.8 seconds. The point source was randomly positioned at distances 1.5--4\,m from the array center and at least 0.5\,m from the walls. To define the microphone positions, a spherical coordinate system, $(r, \theta, \phi)$ was used with its origin at the array center and the azimuthal (xy) plane aligned with the room floor, such that the array’s forward-facing direction corresponded to $(\theta, \phi) = (90^\circ, 0^\circ)$. Within this coordinate system, the microphones were arranged at $r_m = 10\,\mathrm{cm}$, $\theta_m = 90^\circ$, and $\phi_m = 90^\circ - \frac{180^\circ (m-1)}{M-1}$ for $m = 1, \dots, M$, to have the front-looking direction as shown in Fig.~\hyperref[fig:4]{\ref*{fig:4}}. To simulate varying source directions, a relative DOA angle $\phi_{\mathrm{doa}} \in [0^\circ, 60^\circ]$ was randomly selected to place the source to the right of the array. This was implemented by rotating the array accordingly, to produce the source direction illustrated in Fig.~\hyperref[fig:4]{\ref*{fig:4}}. For simplicity, the DOA range was constrained to sources located to the right of the array and within the array’s frontal region.

The BSM-MagLS algorithm transforms the microphone array signals into binaural signals while accounting for the listener’s head orientation. The listener’s head was initially oriented in the same direction as the array, ensuring alignment between the two (see Fig.~\hyperref[fig:1]{\ref*{fig:1}(a)}). The listener's head in this study was represented by HRTF measured using the Neumann KU100 manikin, taken from the Cologne database~\cite{ref23}, employing a Lebedev sampling scheme with 2702 spatial directions. To simulate the rotation of the listener's head during reproduction (see Fig.~\hyperref[fig:1]{\ref*{fig:1}(b)}–\hyperref[fig:1]{(c)}), a rightward rotation angle $\phi_{\text{rot}} \in [21^\circ, 60^\circ]$ was randomly selected. Only rightward rotations were considered to reduce spatial variability and simplify the training process. The input and target binaural signals for the network were generated by applying the BSM-MagLS algorithm to the microphone array recordings under two configurations. In the input configuration, only the listener’s head was rotated by $\phi_{\text{rot}}$ while the array remained fixed (see Fig.~\hyperref[fig:1]{\ref*{fig:1}(b)}). In the target configuration, both the listener’s head and the array were rotated by $\phi_{\text{rot}}$, maintaining alignment between them (see Fig.~\hyperref[fig:1]{\ref*{fig:1}(c)}). After this binaural signal generation step, the dataset was split into training (80\%), validation (10\%), and test (10\%) subsets, ensuring that each set includes the corresponding input-target pairs derived from the same simulated scenes.

\subsection{Methodology}
For the STFT computation, we used a Hann window of 1024 samples (64\,ms), an FFT length of 1024, and a hop size of 256 samples (16\,ms), yielding 513 frequency bins for the 16\,kHz sampling rate. We adopted the same configuration for the small version of SpatialNet as in~\cite{ref16}, which was found to be sufficient for our purposes. The model size, determined by the number of frequency bins due to the full-band mapping architecture, amounts to approximately 3.2\,M parameters in the SpatialNet-small configuration. The network was trained for 40 epochs with a batch size of 8 using the Adam optimizer~\cite{ref24} and an initial learning rate of 0.001. The learning rate was reduced by a factor of 0.5 after epoch 30 if the validation loss did not improve for three consecutive epochs.

Two variants of SpatialNet are considered for training, differing in the choice of binaural loss. For both variants, the training objective combined the signal-level loss defined in~(\ref{eq:11}) with the binaural loss terms, as specified in the overall loss formulation in~(\ref{eq:10}). All loss components were scaled based on their running average to ensure balanced contributions during training. The auditory filter--based binaural loss, defined in~(\ref{eq:12}), was used in the training variant referred to as SpatialNet-AUD, with loss weights $(\alpha, \beta, \gamma)$ for the signal-level loss terms and $(\delta, \lambda, \kappa)$ for the binaural loss terms which were both set to $\{1, 1, 0.5\}$. The alternative variant, SpatialNet-STFT, applied the same signal-level loss with equal weights, but used the STFT-based binaural loss, defined in ~(\ref{eq:28}), with weights $\delta'$ and $\lambda'$ set to $\{1, 1\}$. 
The weight values for both variants were selected to optimize the binaural metric from~\cite{ref17} on the validation set while maintaining signal-level performance.

Performance was evaluated on 3200 speech utterances from the test set for the BSM-MagLS, SpatialNet-STFT, and SpatialNet-AUD methods using both signal-level and binaural metrics to assess reconstruction quality and spatial cue preservation. Each metric was computed using the target binaural signal defined in Section~\ref{sec:3.1} and the corresponding estimated binaural signal produced by each method. The signal-level metrics included SI-SDR, STFT loss, and magnitude-based STFT loss. During evaluation, SI-SDR was computed as in~(\ref{eq:13}), but with equal weighting for the left and right ears, unlike during training. The binaural metrics comprised the auditory filter-based components ILD, IPD, and IVS, as defined in (\ref{eq:21})--(\ref{eq:23}), as well as the perceptual binaural quality metric proposed in~\cite{ref17}, which yields scores between 0 and 100.

\subsection{Results}
Table~\ref{tab:1} presents the evaluation results for SpatialNet-AUD and SpatialNet-STFT, alongside the results for BSM-MagLS that serve as input to both models. Across all evaluated metrics, the trained networks consistently outperform the unprocessed BSM-MagLS signals. In terms of signal-level performance, SpatialNet-AUD and SpatialNet-STFT achieve comparable results, both demonstrating substantial improvement over the input. For the binaural metrics, SpatialNet-AUD exhibits higher accuracy, particularly in ILD, while IPD and IVS scores remain largely similar between the two models.

To assess asymmetries introduced by head rotation in the input BSM-MagLS signals, Table~\ref{tab:2} reports the signal-level metrics separately for the left and right ears. The right ear, positioned farther from the microphone array, consistently shows greater degradation in signal quality. This highlights the importance of accurately reconstructing the far ear signal and demonstrates the effectiveness of the models in correcting these imbalances.

Table~\ref{tab:3} shows signal-level and binaural metric results across increasing head rotation ranges for both SpatialNet-AUD and SpatialNet-STFT, with SpatialNet-STFT values shown in parentheses. For both variants, signal-level performance remains relatively stable across rotation ranges, with a noticeable drop in accuracy appearing only at the highest range. In contrast, binaural metrics exhibit a consistent and progressively worsening trend as head rotation increases, highlighting the challenges in maintaining spatial cue fidelity under larger rotations. Comparing the two variants, signal-level metrics are largely similar, while differences in binaural metrics are most pronounced for ILD, where SpatialNet-AUD achieves substantially higher accuracy. Improvements in IPD and IVS by SpatialNet-AUD over SpatialNet-STFT are smaller but consistently evident.

\begin{table}[!t]
\centering
\setlength{\tabcolsep}{3pt} 
\caption{Comparison of Signal-Level and Binaural Metrics for BSM-MagLS, SpatialNet-STFT, and SpatialNet-AUD}
\label{tab:1}
\resizebox{\linewidth}{!}{%
\begin{tabular}{l@{\hskip 2pt}|c@{\hskip 2pt}c@{\hskip 2pt}c@{\hskip 2pt}|c@{\hskip 2pt}c@{\hskip 2pt}c@{\hskip 2pt}}
\toprule
\multicolumn{1}{c|}{} &
\multicolumn{3}{c|}{\textbf{Signal-Level Metrics}} & 
\multicolumn{3}{c}{\textbf{Binaural Metrics}} \\
\textbf{Method} & ↑ SI-SDR [dB] & ↓ $\mathcal{L}_{\text{STFT}}$ & ↓ $\mathcal{L}_{\text{Mag-STFT}}$ & ↓ $\mathcal{L}_{\text{ILD}}$ [dB] & ↓ $\mathcal{L}_{\text{IPD}}$ [rad] & ↓ $\mathcal{L}_{\text{IVS}}$ \\
\midrule
BSM-MagLS & 5.17 & 0.351 & 1.17 & 2.04 & 0.59 & 0.046 \\
SpatialNet-STFT & 8.79 & 0.293 & 0.99 & 1.77 & 0.51 & 0.044 \\
SpatialNet-AUD & \textbf{8.86} & \textbf{0.291} & \textbf{0.96} & \textbf{1.51} & \textbf{0.48} & \textbf{0.042} \\
\bottomrule
\end{tabular}%
}
\end{table}

\begin{table}[!t]
\centering
\caption{Signal-Level Metrics for BSM-MagLS Outputs, Reported Separately for the Left and Right Ears}
\label{tab:2}
\resizebox{\linewidth}{!}{%
\begin{tabular}{l|ccc}
\toprule
\textbf{Method} & ↑ SI-SDR [dB] & ↓ $\mathcal{L}_{\text{STFT}}$ & ↓ $\mathcal{L}_{\text{Mag-STFT}}$ \\
\midrule
BSM-MagLS (right ear) & 1.87 & 0.50 & 1.29 \\
BSM-MagLS (left ear) & \textbf{8.47} & \textbf{0.19} & \textbf{1.05} \\
\bottomrule
\end{tabular}%
}
\end{table}

Finally, Table~\ref{tab:4} reports binaural quality scores, computed using the perceptual metric from~\cite{ref17}, across increasing head rotation ranges for BSM-MagLS, SpatialNet-STFT, and SpatialNet-AUD, comparing them with the target signal. The binaural scores of BSM-MagLS decline noticeably with increasing head rotation, underscoring the growing need to correct its outputs under such conditions, particularly at larger rotation angles. This trend reflects the increased degradation in the ear farther from the array, leading to reduced spatial fidelity in the reconstructed signal~\cite{ref5}. SpatialNet-AUD performs robustly across all head rotation ranges. In comparison, SpatialNet-STFT exhibits a more substantial decline in performance as head rotation increases. These results highlight the importance of the proposed auditory filter-based binaural loss in preserving spatial cues and correcting BSM-MagLS degradation under head rotation.

\section{Listening Experiment}

This section presents a listening experiment conducted to subjectively evaluate the quality of the binaural signals produced by the methods considered in this study. It complements the previous simulation study in Section~\ref{sec:5}, which relied solely on objective and perceptually motivated metrics.

For this purpose, two distinct acoustic environments were simulated using the image source method, implemented in MATLAB. In both environments, a three-second female speech excerpt from the TSP database~\cite{ref25}, resampled to 16~kHz, served as the source signal. The first environment, env1, consisted of a $10 \times 6 \times 3$~m room with a reverberation time of $T_{60} = 0.4$~s, where a point source was placed at position $(7.4, 4.8, 1.7)$~m in the room, and the center of the six-microphone semi-circular array at $(6.7, 1.5, 1.7)$~m. The second environment, env2, was a $7 \times 8 \times 3.5$~m room with $T_{60} = 0.68$~s, with the source located at $(4.2, 5.9, 1.7)$~m and the array centered at $(5.4, 3.8, 1.7)$~m. In both cases, the microphone signals were computed according to~(1), without additive noise. For each of the two acoustic environments, BSM-MagLS filters were computed for head rotations of $\phi_{\text{rot}} = 60^\circ, 90^\circ$, following the input and target configurations defined in Section~\ref{sec:5.1}. In the target configuration, both the listener’s head and the microphone array were rotated by $\phi_{\text{rot}}$, maintaining alignment between them. In the input configuration, only the listener’s head was rotated while the array remained fixed, and the compensated BSM-MagLS outputs were subsequently processed by SpatialNet-AUD and SpatialNet-STFT to produce the estimated binaural signals. Binaural quality scores, obtained using the perceptual binaural quality metric proposed in~\cite{ref17} and computed relative to the target, are reported in Table~\ref{tab:5} as an objective reference for the specific conditions considered in the listening experiment.

\begin{table}[!t]
\centering
\setlength{\tabcolsep}{3pt} 
\caption{Signal-Level and Binaural Metrics Across Increasing Head Rotation Ranges, Reported for SpatialNet-AUD (SpatialNet-STFT Values in Subscripts)}
\label{tab:3}
\resizebox{\linewidth}{!}{%
\begin{tabular}{c@{\hskip 2pt}|c@{\hskip 2pt}c@{\hskip 2pt}c@{\hskip 2pt}|c@{\hskip 2pt}c@{\hskip 2pt}c@{\hskip 2pt}}
\toprule
\multicolumn{1}{c|}{} &
\multicolumn{3}{c|}{\textbf{Signal-Level Metrics}} & 
\multicolumn{3}{c}{\textbf{Binaural Metrics}} \\
\textbf{Rotation Range} & ↑ SI-SDR [dB] & ↓ $\mathcal{L}_{\text{STFT}}$ & ↓ $\mathcal{L}_{\text{Mag-STFT}}$ & ↓ $\mathcal{L}_{\text{ILD}}$ [dB] & ↓ $\mathcal{L}_{\text{IPD}}$ [rad] & ↓ $\mathcal{L}_{\text{IVS}}$ \\
\midrule
21--30 & 9.23$_{(9.14)}$ & \textbf{0.22}$_{(0.22)}$ & 0.99$_{(1.02)}$ & \textbf{1.29}$_{(1.39)}$ & \textbf{0.34}$_{(0.36)}$ & \textbf{0.034}$_{(0.036)}$ \\
31--40 & 9.86$_{(9.86)}$ & 0.29$_{(0.29)}$ & \textbf{0.90}$_{(0.93)}$ & 1.39$_{(1.63)}$ & 0.43$_{(0.46)}$ & 0.039$_{(0.041)}$ \\
41--50 & \textbf{9.91}$_{(9.75)}$ & 0.28$_{(0.29)}$ & 0.92$_{(0.95)}$ & 1.56$_{(1.87)}$ & 0.50$_{(0.55)}$ & 0.043$_{(0.046)}$ \\
51--60 & 6.69$_{(6.60)}$ & 0.35$_{(0.35)}$ & 1.03$_{(1.06)}$ & 1.79$_{(2.16)}$ & 0.64$_{(0.69)}$ & 0.050$_{(0.053)}$ \\
\bottomrule
\end{tabular}%
}
\end{table}

\begin{table}[!t]
\centering
\caption{Binaural Quality Scores Across Head Rotation Ranges for BSM-MagLS, SpatialNet-STFT, and SpatialNet-AUD}
\label{tab:4}
\resizebox{\linewidth}{!}{%
\begin{tabular}{c|ccc}
\toprule
\textbf{Rotation Range} & \textbf{BSM-MagLS} & \textbf{SpatialNet-STFT} & \textbf{SpatialNet-AUD} \\
\midrule
21--60 & 75.68 & 80.23 & \textbf{84.82} \\
\cmidrule(lr){1-4}
21--30 & 86.61 & 88.65 & \textbf{89.80} \\
31--40 & 79.86 & 84.05 & \textbf{86.92} \\
41--50 & 72.92 & 77.08 & \textbf{83.65} \\
51--60 & 65.55 & 70.92 & \textbf{79.49} \\
\bottomrule
\end{tabular}%
}
\end{table}

The MUltiple Stimuli with Hidden Reference and Anchor (MUSHRA) test~\cite{ref26} was employed for the subjective evaluation. Four MUSHRA screens were generated, one for each combination of acoustic environment and head rotation ($\phi_{\text{rot}} = 60^\circ$ and $90^\circ$). In each screen, the reference was taken to be the binaural signal of the target configuration, and the anchor was a first-order Ambisonics (FOA) signal. Hence, the five test signals in each screen comprised the hidden reference, the FOA anchor, the compensated BSM-MagLS, and the estimations obtained with SpatialNet-AUD and SpatialNet-STFT. The MUSHRA screens and signals were presented in a randomized order for each subject. The scoring criterion for evaluating the similarity between the test signals and the reference was defined as overall quality, described to the subjects in terms of both spatial and timbral variations. Ratings were given on a scale from 0 to 100, where 100 indicated that the test signal was indistinguishable from the reference. Ten subjects with no known hearing impairments participated in the experiment. Prior to the listening test, a training stage was conducted with a single screen to familiarize the subjects with the scoring procedure. Informed consent was obtained from all participants prior to the experiment.

\begin{figure}[!t] 
\centering
\includegraphics[width=\linewidth]{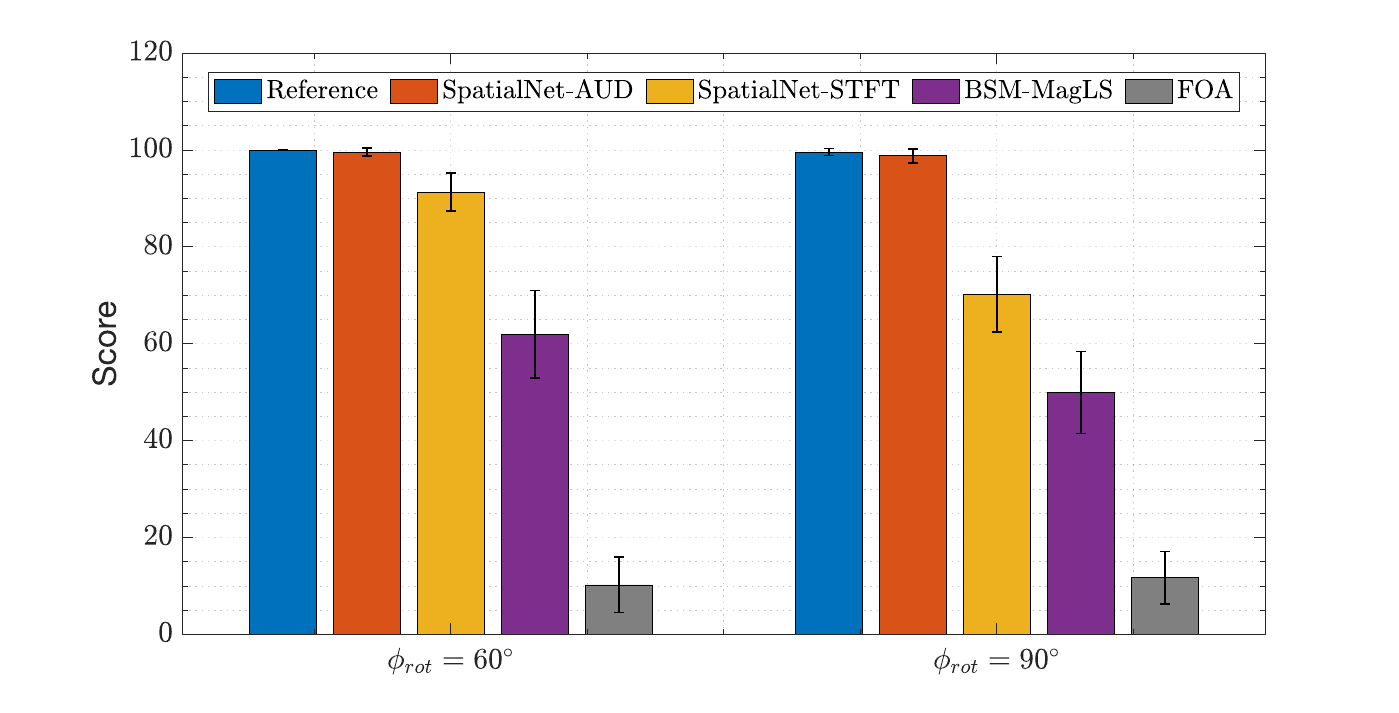}
\vskip -2ex 
\caption{Estimated marginal means of MUSHRA scores across head rotations ($\phi_{\text{rot}}$), averaged over the two acoustic environments. Bars show marginal means and error bars indicate 95\% confidence intervals.}
\label{fig:5}
\end{figure}

The scores assigned by the subjects were analyzed using a repeated-measures ANOVA with three within-subject factors: (a) the head rotation angle ($\phi_{\text{rot}} = 60^{\circ}$ and $90^{\circ}$), (b) the binaural reproduction method (Reference, BSM-MagLS, SpatialNet-AUD, SpatialNet-STFT, and FOA), and (c) the acoustic environment (env1 and env2). The analysis revealed significant main effects of head rotation angle, $F(1, 180) = 21.1$, $p < .001$, $\eta^2 = 0.11$, and reproduction method, $F(4, 180) = 496.8$, $p < .001$, $\eta^2 = 0.92$. In contrast, the effect of environment was not significant, $F(1, 180) = 0.18$, $p = 0.68$, despite differences in room dimensions, reverberation, and source and array positions. Therefore, the means and 95\% confidence intervals of the scores given to each test signal, presented as boxplots in Fig.~\hyperref[fig:5]{\ref*{fig:5}}, were averaged across environments.

\begin{table}[!t]
\centering
\caption{Binaural Quality Scores for the Test Signals Used in the Listening Experiment, Obtained with BSM-MagLS, SpatialNet-STFT, and SpatialNet-AUD, Using the Perceptual Metric From~\cite{ref17}}
\label{tab:5}
\resizebox{\linewidth}{!}{%
\begin{tabular}{c|ccc}
\toprule
\textbf{Condition} & \textbf{BSM-MagLS} & \textbf{SpatialNet-STFT} & \textbf{SpatialNet-AUD} \\
\midrule
env1 - $\phi_{\text{rot}} = 60^\circ$ & 63.26 & 66.21 & \textbf{75.04} \\
env1 - $\phi_{\text{rot}} = 90^\circ$ & 53.96 & 57.87 & \textbf{71.16} \\
env2 - $\phi_{\text{rot}} = 60^\circ$ & 65.28 & 68.57 & \textbf{77.33} \\
env2 - $\phi_{\text{rot}} = 90^\circ$ & 56.19 & 60.68 & \textbf{71.45} \\
\bottomrule
\end{tabular}%
}
\end{table}

A significant interaction between head rotation and reproduction method, $F(4, 180) = 10.3$, $p < .001$, $\eta^2 = 0.19$, indicated that the effect of head orientation differed among the reproduction methods. No statistically significant interactions were found involving the environment factor. Since the interaction between head rotation and reproduction method was statistically significant, a post-hoc analysis with Bonferroni correction was conducted for this interaction.

The interaction was examined first for a fixed degree of head rotation. For a head rotation of $\phi_{\text{rot}} = 60^{\circ}$, SpatialNet-AUD achieved mean scores only 0.5 points below the reference, a difference that was not statistically significant ($p = .887$), indicating perceptual indistinguishability from the target. Furthermore, the mean score of SpatialNet-AUD exceeded that of BSM-MagLS and SpatialNet-STFT by 44.5 and 10.35 points, respectively, with both differences reaching statistical significance ($p < .001$ and $p = .002$). At $\phi_{\text{rot}} = 90^{\circ}$, SpatialNet-AUD again maintained high perceptual quality, with a mean difference of 0.8 points from the reference, which was not statistically significant ($p = .821$). Compared to SpatialNet-AUD, BSM-MagLS and SpatialNet-STFT achieved substantially lower scores at $90^{\circ}$, by 54.5 and 35.2 points respectively, and both differences were statistically significant ($p < .001$). Examining the relative performance between $60^{\circ}$ and $90^{\circ}$, BSM-MagLS showed a modest decline of 10.3 points, whereas SpatialNet-STFT exhibited a considerably larger decrease of 25.15 points. FOA consistently received the lowest ratings across both head rotations, with mean scores significantly below all other methods ($p < .001$).

The interaction was then examined for a fixed binaural reproduction method. In this analysis, the differences between the mean scores of BSM-MagLS and of SpatialNet-STFT across $\phi_{\text{rot}} = 60^{\circ}$ and $\phi_{\text{rot}} = 90^{\circ}$ were both statistically significant ($p = .003$ and $p < .001$, respectively), confirming the sensitivity of these methods to head rotation. By contrast, similar comparisons for the remaining reproduction methods (Reference, SpatialNet-AUD, FOA) did not yield statistically significant differences, indicating perceptual stability for those signals under the examined rotations.

These results indicate that under substantial head rotations, the SpatialNet-AUD method may produce binaural signals that are significantly better than those generated by BSM-MagLS, whose performance further declined for the larger head rotation angle. Although both SpatialNet models were trained on head rotations within $\phi_{\text{rot}} \in [21^{\circ}, 60^{\circ}]$, they outperformed BSM-MagLS at $\phi_{\text{rot}} = 90^{\circ}$, indicating their capacity to generalize beyond the training range. SpatialNet-AUD maintained performance levels nearly indistinguishable from the reference and remained robust across both rotations, while SpatialNet-STFT exhibited a pronounced decline in performance between $60^{\circ}$ and $90^{\circ}$, reflecting a reduced ability to generalize to larger head rotations compared with SpatialNet-AUD.

\section{Conclusions}
In this work, we studied the use of SpatialNet as a post-processing framework for correcting BSM-MagLS outputs under head rotations. While the approach of BSM-MagLS to head-rotation compensation achieves accurate reproduction for small rotations, its performance degrades as the head turns farther from the array. The proposed post-processing mitigates these degradations, reducing distortions and yielding more consistent reproduction across rotations. Subjective evaluation confirmed that employing an auditory filter–based binaural loss yields superior perceived quality compared with binaural losses used in earlier studies. Furthermore, the method can achieve reproduction quality comparable to the reference condition, obtained with BSM-MagLS when the array is aligned with the head, while remaining robust as rotation increases. These findings highlight the potential of integrating deep learning with BSM-MagLS to enable robust, high-quality binaural reproduction with arbitrary microphone arrays under head-tracking conditions. It is suggested that future work extends the method to diverse HRTFs, broader and bidirectional head-rotation scenarios, more complex real-world acoustic environments, and arbitrary microphone configurations. Additional directions include employing a more task-specific neural architecture and an extension of the listening test performed in this work.


\begin{thebibliography}{99}
\bibliographystyle{IEEEtran}

\bibitem{ref1}
B. Rafaely, V. Tourbabin, E. Habets, Z. Ben-Hur, H. Lee, H. Gamper, L. Arbel, L. Birnie, T. Abhayapala, and P. Samarasinghe,
``Spatial audio signal processing for binaural reproduction of recorded acoustic scenes–review and challenges,''
\textit{Acta Acust.}, vol. 6, p. 47, 2022.

\bibitem{ref2}
M. A. Gerzon, ``Ambisonics in multichannel broadcasting and video,''
\textit{J. Audio Eng. Soc.}, vol. 33, no. 11, pp. 859--871, 1985.

\bibitem{ref3}
L. Madmoni \textit{et al.},
``Design and analysis of binaural signal matching with arbitrary microphone arrays and listener head rotations,''
\textit{EURASIP Journal on Audio, Speech, and Music Processing}, vol. 2025, no. 1, p. 11, 2025.


\bibitem{ref4}
C. Schorkhuber, M. Zaunschirm, and R. Höldrich,
``Binaural rendering of ambisonic signals via magnitude least squares,''
in \textit{Proc. DAGA}, vol. 44, 2018, pp. 339--342.

\bibitem{ref5}
L. Madmoni \textit{et al.},
``Binaural reproduction from microphone array signals incorporating head-tracking,''
in \textit{Proc. I3DA}, 2021.

\bibitem{ref6}
L. Mccormack \textit{et al.},
``Six-degrees-of-freedom binaural reproduction of head-worn microphone array capture,''
\textit{J. Audio Eng. Soc.}, vol. 71, no. 10, pp. 638--649, 2023.

\bibitem{ref7}
A. Berger \textit{et al.},
``Insights into the incorporation of signal information in binaural signal matching with wearable microphone arrays,''
\textit{arXiv:2409.11731}, 2024.

\bibitem{ref8}
V. Tokala, E. Grinstein, M. Brookes, S. Doclo, J. Jensen, and P. A. Naylor,
``Binaural speech enhancement using deep complex convolutional transformer networks,''
in \textit{Proc. ICASSP}, 2024, pp. 681--685.

\bibitem{ref9}
C. Hernandez-Olivan \textit{et al.},
``Interaural time difference loss for binaural target sound extraction,''
in \textit{Proc. IWAENC}, 2024.

\bibitem{ref10}
M. Dietz, S. D. Ewert, and V. Hohmann,
``Auditory model-based direction estimation of concurrent speakers from binaural signals,''
\textit{Speech Commun.}, vol. 53, no. 5, pp. 592--605, 2011.

\bibitem{ref11}
H. L. Van Trees,
\textit{Optimum Array Processing: Part IV of Detection, Estimation, and Modulation Theory}. Hoboken, NJ, USA: Wiley, 2004.

\bibitem{ref12}
L. Madmoni, J. Donley, V. Tourbabin, and B. Rafaely,
``Beamforming-based binaural reproduction by matching of binaural signals,''
in \textit{AES Conf.: Audio for Virtual and Augmented Reality}, 2020.

\bibitem{ref13}
A. N. Tikhonov, A. Goncharsky, V. Stepanov, and A. G. Yagola,
\textit{Numerical Methods for the Solution of Ill-Posed Problems}.
Berlin, Germany: Springer, 2013, vol. 328.

\bibitem{ref14}
E. A. Macpherson and J. C. Middlebrooks,
``Listener weighting of cues for lateral angle: The duplex theory of sound localization revisited,''
\textit{J. Acoust. Soc. Am.}, vol. 111, no. 5, p. 2219, 2002.

\bibitem{ref15}
R. Klumpp and H. Eady,
``Some measurements of interaural time difference thresholds,''
\textit{J. Acoust. Soc. Am.}, vol. 28, no. 5, pp. 859--860, 1956.

\bibitem{ref16}
C. Quan and X. Li,
``SpatialNet: Extensively learning spatial information for multichannel joint speech separation, denoising and dereverberation,''
\textit{IEEE/ACM Trans. Audio Speech Lang. Process.}, vol. 32, pp. 1310--1323, 2024.

\bibitem{ref17}
T. Biberger, H. Schepker, F. Denk, and S. D. Ewert,
``Instrumental quality predictions and analysis of auditory cues for algorithms in modern headphone technology,''
\textit{Trends Hear.}, vol. 25, p. 23312165211001219, 2021.

\bibitem{ref18}
J. Le Roux \textit{et al.},
``SDR–half-baked or well done?,'' in \textit{Proc. ICASSP}, 2019.

\bibitem{ref19}
G. F. Kuhn,
``Model for the interaural time differences in the azimuthal plane,''
\textit{J. Acoust. Soc. Am.}, vol. 62, no. 1, pp. 157--167, 1977.

\bibitem{ref20}
K. Ito, ``The LJ Speech dataset.'' [Online]. Available: https://keithito.com/LJ-Speech-Dataset/, 2017.

\bibitem{ref21}
J. B. Allen and D. A. Berkley,
``Image method for efficiently simulating small-room acoustics,''
\textit{J. Acoust. Soc. Am.}, vol. 65, no. 4, pp. 943--950, 1979.

\bibitem{ref22}
MATLAB, version 9.10.0 (R2021a). Natick, MA, USA: The MathWorks Inc., 2021.

\bibitem{ref23}
B. Bernschütz,
``A spherical far field HRIR/HRTF compilation of the Neumann KU 100,''
in \textit{Proc. AIA/DAGA}, 2013, p. 29.

\bibitem{ref24}
D. P. Kingma and J. L. Ba,
``Adam: A method for stochastic optimization,''
in \textit{Proc. ICLR}, 2015.

\bibitem{ref25}
P. Kabal, ``TSP speech database,'' \textit{Tech. Rep.}, 2002.

\bibitem{ref26}
ITU Recommendation,
``Method for the subjective assessment of intermediate quality level of coding systems,''
\textit{ITU-R BS}, pp. 1534--1, 2003.

\end{thebibliography}
\end{document}